\documentclass[aps,pra,nofootinbib,amsfonts,amssymb,amsmath]{revtex4}

\begin{document}

\title{Prepotential approach to quasinormal modes}

\author{Choon-Lin Ho}
\affiliation{Department of Physics, Tamkang University, Tamsui
251, Taiwan, R.O.C.}

%\date{Oct 15, 2010}

\begin{abstract}

In this paper we demonstrate how the recently reported exactly and
quasi-exactly solvable models admitting quasinormal modes can be
constructed and classified very simply and directly by the newly
proposed prepotential approach.  These new models were previously
obtained within the Lie-algebraic approach.  Unlike the
Lie-algebraic approach, the prepotential approach does not require
any knowledge of the underlying symmetry of the system.  It treats
both quasi-exact and exact solvabilities on the same footing, and
gives the potential as well as the eigenfunctions and eigenvalues
simultaneously. We also present three new models with quasinormal
modes: a new exactly solvable Morse-like model, and two new
quasi-exactly solvable models of the Scarf II and generalized
P\"oschl-Teller types.

\end{abstract}

%(Formalism, Sol'n to wave eqns, Scattering theory, Algebraic methods)

\maketitle

\section{Introduction}

Quasinormal modes (QNM) has attracted great interest in recent
years \cite{QNM}.  They arise as waves emitted by a perturbed
neutron star or black hole that are outgoing to spatial infinity
and the event horizon.   Generally, the wave function of QNMs has
discrete complex frequency, whose imaginary part leads to a
damping behavior.  QNM carry information of black holes and
neutron stars, and thus are of importance to gravitational-wave
astronomy. In fact, these oscillations, produced mainly during the
formation phase of the compact stellar objects, can be strong
enough to be detected by several large gravitational wave
detectors under construction.

As black hole potentials are generally too complicated to allow
analytic treatment, so in order to understand the origin of the
discrete imaginary frequencies, one can try approximating the top
region of the black hole potential by some inverted potentials
which are solvable.  This has been done by using the inverted
harmonic oscillator \cite{Kim},  and the P\"oschl-Teller potential
\cite{FM}.

Recently, in \cite{ChoHo1} we have extended the number of exactly
solvable models that admit QNMs.  We take QNMs to include both
decaying and growing modes with complex energies. Furthermore, we
have provided the first model with QNM that is quasi-exactly
solvable (QES).  A system is called QES if a part of its spectrum,
but not the whole spectrum, can be determined analytically
\cite{TU,Tur,Tur1,GKO,Shifman,Ush,PT,KMO}. Our approach in
\cite{ChoHo1} was to study solutions of QNM based on the
$sl(2)$-Lie-algebraic approach to one-dimensional QES theory
\cite{TU,Tur,GKO,Shifman}. We demonstrated that, by suitably
complexifying some parameters of the generators of the $sl(2)$
algebra while keeping the Hamiltonian Hermitian, we could indeed
obtain potentials admitting exact or quasi-exact QNMs. These
models were later re-studied numerically by the asymptotic
iteration method in \cite{OR}.

In this paper we would like to show that exactly solvable and QES
models with QNMs can be constructed much more simply without
resorting to the machinery of Lie-algebra.  This is achieved
through the prepotential approach proposed recently
\cite{Ho1,Ho2,Ho3,Ho4}. This is a simple constructive approach,
based on the so-called prepotential \cite{Tur1,ST,Ho5}, which can
give the potential as well as the eigenfunctions and eigenvalues
simultaneously.  The novel feature of the approach is that both
exact and quasi-exact solvabilities can be solely classified by
two integers, the degrees of two polynomials which determine the
change of variables and the zero-th order prepotential.  Hence
this approach treats both quasi-exact and exact solvabilities on
the same footing, and it provides a simple way to determine the
required change of variables, say $x$, to a new one $z=z(x)$. All
the well-known exactly solvable models classified in
supersymmetric quantum mechanics (SUSYQM) \cite{Cooper}, the QES
models discussed in \cite{TU,Tur,GKO,Shifman,Ush}, and some new
QES ones (also for non-Hermitian Hamiltonians), can be generated
by appropriately choosing the two polynomials. Our approach,
unlike the Lie-algebraic approach, does not require any knowledge
of the underlying symmetry of the system. Furthermore, our
approach can generate the Coulomb, Eckart, Rosen-Morse type I and
II models \cite{Ho3} which are not covered by the standard
Lie-algebraic program \cite{GKO,Shifman}. Compared with SUSYQM,
our approach has the advantage that we do not have to assume the
sufficient condition for integrability needed in SUSYQM, namely,
shape invariance. In fact, shape invariance comes out
automatically from this approach \cite{Ho4}. What is more, the
transformation of the original variable $z(x)$ is determined
within the prepotential approach, whereas in SUSYQM this has to be
taken as given from the known solutions of the respective models
before one could solve the shape invariance condition.

We shall adopt the prepotential approach here to generate and
classify all one-dimensional exactly solvable and QES models with
QNMs based on sinusoidal coordinates, namely, those coordinates
$z(x)$ whose derivatives (with respect to $x$) squared are at most
quadratic in $z$. Our strategy is to suitably complexify some or
all of the parameters in the prepotential while keeping the
resulted potentials real. We find that all the systems reported in
\cite{ChoHo1} can be very easily constructed.  During the course
of investigation, we also realize that there are three new QNM
models which were missed in \cite{ChoHo1}. We thus take this
opportunity to report on them.

The plan of the paper is as follows.  In Sect.~II we briefly
review the essence of the prepotential approach. Sect.~III to V
then discuss the prepotential construction of QNM models of the
Scarf II, the Morse, and the generalized P\"oschl-Teller type,
respectively. These three types of potentials have some
interesting features, and serve as good examples to illustrate the
procedure. Furthermore, there is one new QNM model in each of
these three types that we would like to report. Sect.~VI
summarizes the paper. In Appendices A to C, we list all other
cases for easy reference.

\section{Prepotential approach}

The main ideas of the prepotential approach \cite{Ho1,Ho2} can be
summarized as follows (we adopt the unit system in which $\hbar$
and the mass $m$ of the particle are such that $\hbar=2m=1$).
Consider a wave function $\phi_N(x)$ ($N$: non-negative integer)
which is defined as
\begin{eqnarray}
\phi_N(x)\equiv e^{-W_0(x)}p_N (z),\label{phi_N}
\end{eqnarray}
with
\begin{eqnarray}
p_N (z)\equiv \left\{
               \begin{array}{ll}
                1, & N=0;\\
                \prod_{k=1}^N (z-z_k),& N>0.
                        \end{array}
               \right.
\label{p_N}
\end{eqnarray}
Here $z=z(x)$ is some real function of the basic variable $x$,
$W_0(x)$ is a regular function of $z(x)$, and $z_k$'s are the
roots of $p_N(z)$. The variable $x$ is defined on the full line,
half-line, or finite interval, as dictated by the choice of
$z(x)$. The function $p_N(z)$ is a polynomial in an
$(N+1)$-dimensional Hilbert space with the basis $\langle
1,z,z^2,\ldots,z^N \rangle$. $W_0(x)$ defines the ground state
wave function.

The wave function $\phi_N$ can be recast as
\begin{eqnarray}
 \phi_N =\exp\left(- W_N(x,\{z_k\})
\right), \label{f2}
\end{eqnarray}
with $W_N$ given by
\begin{eqnarray}
W_N(x,\{z_k\}) = W_0(x) - \sum_{k=1}^N \ln |z(x)-z_k|. \label{W}
\end{eqnarray}
Operating on $\phi_N$ by the operator $-d^2/dx^2$ results in a
Schr\"odinger equation $H_N\phi_N=0$, where
\begin{eqnarray}
H_N &=&-\frac{d^2}{dx^2} + V_N,\\
V_N&\equiv&  W_N^{\prime 2} - W_N^{\prime\prime}.
\end{eqnarray}
Here and below the prime represents derivative with respect to
$x$. Since the potential $V_N$ is determined by $W_N$, we thus
call $W_N$ the $N$th order prepotential.  From Eq.~(\ref{W}), one
finds that $V_N$ has the form $V_N=V_0+\Delta V_N$:
\begin{eqnarray}
V_0 &=&W_0^{\prime 2} - W_0^{\prime\prime},\nonumber\\
 \Delta V_N &=&
-2\left(W_0^\prime z^\prime
-\frac{z^{\prime\prime}}{2}\right)\sum_{k=1}^N \frac{1}{z-z_k} +
\sum_{{k,l}\atop{k\neq l}} \frac{z^{\prime 2}}{(z-z_k)(z-z_l)}.
\label{V}
\end{eqnarray}

Thus the form of $V_N$, and consequently its solvability, are
determined by the choice of $W_0(x)$ and $z^{\prime 2}$ (or
equivalently by $z^{\prime\prime}=(dz^{\prime 2}/dz)/2$). Let
$W_0^\prime z^\prime=P_m(z)$ and $z^{\prime 2}=Q_n(z)$ be two
polynomials of degree $m$ and $n$ in $z$, respectively. The
variables $x$ and $z$ are related by
\begin{eqnarray}
x(z)=\pm \int^z \frac{dz}{\sqrt{Q_n(z)}},\label{z(x)}
\end{eqnarray}
and the prepotential $W_0(x)$ is determined as
\begin{eqnarray}
W_0(x)=\left(\int^z dz
\frac{P_m(z)}{Q_n(z)}\right)_{z=z(x)}.\label{W0}
\end{eqnarray}
We assume (\ref{z(x)}) is invertible to give $z=z(x)$.
Eqs.~(\ref{z(x)}) and (\ref{W0}) define the change of variables
$z(x)$ and the corresponding prepotential $W_0(x)$. Thus, $P_m(z)$
and $Q_n(z)$ determine the quantum system.  Of course, for bound
state problems the choice of $P_m$ and $Q_n$ must ensure
normalizability of $\phi_0=\exp(-W_0)$.

Now depending on the degrees of the polynomials $P_m$ and $Q_n$,
we have the following situations \cite{Ho1,Ho2}:

\begin{enumerate}
\item[(i)] if $\max\{m,n-1\}\leq 1$, then in $V_N(x)$
the parameter $N$ and the roots $z_k$'s will only appear as an
additive constant and not in any term involving powers of $z$.
Such system is then exactly solvable;

\item[(ii)] if $\max\{m,n-1\}=2$, then $N$ may appear in
the first power term in $z$, but $z_k$'s only in an additive term.
If $N$ does appear before the $z$-term, then the system  belongs
to the so-called type 1 QES system defined in \cite{Tur}, i.e.,
for each $N\geq 0$, $V_N$ admits $N+1$ solvable states with the
eigenvalues being given by the $N+1$ sets of roots $z_k$'s.  This
is the main type of QES systems considered in the literature;

\item[(iii)] if $\max\{m,n-1\}\geq 3$
\footnote{We take this opportunity to correct a typographic error
in \cite{Ho1,Ho2}, where the condition for $m$ and $n$ for this
case was erroneously written as $\min\{m,n-1\}\geq 3$.}, then not
only $N$ but also $z_k$'s may appear in terms involving powers of
$z$. If $z_k$'s do appear before any $z$-dependent term, then for
each $N\geq 0$, there are $N+1$ different potentials $V_N$,
differing in several parameters in terms involving powers of $z$,
have the same eigenvalue (when the additive constant, or the zero
point, is appropriately adjusted). When $z_k$'s appear only in the
first power term in $z$, such systems are called type 2 QES
systems in \cite{Tur}.  We see that QES models of higher types are
possible.
\end{enumerate}

This gives a very simple algebraic classification of exact and
quasi-exact solvabilities. Previously exact and QES systems were
treated separately.

In the rest of this paper we shall consider only cases with
$m,~n\leq 2$. Coordinates with $n\leq 2$ are called the sinusoidal
coordinates. Let $P_2(z)=A_2 z^2 + A_1 z + A_0$ and $Q_2(z)=\alpha
z^2 + \beta z + \gamma$. Hence the solvability of the system is
determined solely by $A_2$: exactly solvable if $A_2=0$, or (type
1) QES otherwise. The potential $V_N$ takes the form

\begin{eqnarray}
 V_N ={W_0^\prime}^2 - W_0^{\prime\prime}+ \alpha N^2 -2 A_1N
 - 2 A_2 N z - 2A_2\sum_{k=1}^N z_k
 -2\sum_{k=1}^N
\frac{1}{z-z_k}\left\{P_2(z_k)-\frac{\alpha}{2}z_k -
\frac{\beta}{4} - \sum_{l\neq k} \frac{Q_2(z_k)}{z_k-z_l}\right\}.
\label{V-1}
\end{eqnarray}
Demanding the residues at $z_k$'s vanish gives the Bethe ansatz
equations satisfied by the roots $z_k$'s:
\begin{eqnarray}
P_2(z_k)-\frac{\alpha}{2}z_k - \frac{\beta}{4} - \sum_{l\neq k}
\frac{Q_2(z_k)}{z_k-z_l}=0,~~k=1,2,\ldots,N,
\end{eqnarray}
or
\begin{equation}
A_2 z_k^2 + \left(A_1 - \frac{\alpha}{2}\right) z_k +
A_0-\frac{\beta}{4}-\sum_{l\neq k}\frac{\alpha z_k^2 + \beta z_k
+\gamma}{z_k-z_l}=0.\label{BAE}
\end{equation}

Using
\begin{eqnarray}
W_0^\prime (z)=
\frac{P_2(z)}{\sqrt{Q_2(z)}},~~W_0^{\prime\prime}(z)
=z^\prime\frac{dW_0^\prime}{dz}=\frac{Q_2 \frac{dP_2}{dz} -
\frac12 P_2 \frac{dQ_2}{dz}}{Q_2},
\end{eqnarray}
we arrive at the potential
\begin{eqnarray}
V_N (x) &=& \frac{P_2^2 - Q_2 \frac{dP_2}{dz} + \frac12 P_2
\frac{dQ_2}{dz}}{Q_2}-\left(2 A_1N - \alpha N^2
 +2 A_2 N z + 2A_2\sum_{k=1}^N z_k\right)\nonumber\\
&=&\left[\frac{(A_2 z^2 + A_1 z + A_0)^2}{\alpha z^2 + \beta z +
\gamma} -2(N+1)A_2z  +\frac12\left(A_2 z^2 + A_1 z +
A_0\right)\frac{2\alpha z + \beta}{\alpha z^2 + \beta z +
\gamma}\right]_{z=z(x)}\nonumber\\
&& -\left[(2N +1)A_1- \alpha N^2
  + 2A_2\sum_{k=1}^N z_k\right],
 \label{V-2}
\end{eqnarray}
and the wave function
\begin{eqnarray}
\psi_N &\sim& e^{-W_0}p_N(z)\nonumber\\
&\sim& e^{\int^{z(x)} dz
\frac{P_2(z)}{Q_2(z)}}\,p_N(z).\label{wf-N}
\end{eqnarray}
Eq.~(\ref{V-2}) gives the most general form of potential, based on
sinusoidal coordinates, that cover both the exactly and
quasi-exactly solvable systems.

It turns out that there are only three inequivalent canonical
forms of the sinusoidal coordinates \cite{Ho4}, namely, (i)
$z^{\prime 2}=\gamma\neq 0$, (ii) $z^{\prime 2}=\beta z$ ($\beta>
0$), and (iii) $z^{\prime 2}=\alpha (z^2  + \delta)$
($\delta=0,\pm 1$ for $\alpha>0$, and $\delta=-1$ if $\alpha<0$).
Case (i) and (ii) correspond to one- and three-dimensional
oscillator-type potentials, respectively. In case (iii), for
$\alpha>0$, the potentials are of the Scarf II ($\delta=1$), Morse
($\delta=0$) and generalized P\"oschl-Teller ($\delta=-1$) types,
while for $\alpha<0$ and $\delta=-1$, the potentials generated
belong to the Scarf I type.

For clarity of presentation, in the main text we shall illustrate
the prepotential construction of QNM models only for the Scarf II,
Morse, and generalized P\"oschl-Teller type potentials. Other
cases are summarized in the Appendices.  As mentioned in Sect.~I,
we choose to discuss these three cases because they have some
interesting features that serve as good examples to illustrate the
procedure, and because there are new QNM models in these types not
realized in \cite{ChoHo1}.

\section{Scarf II: $z^{\prime 2}=\alpha (z^2  + 1)$}

Consider first the case $z^{\prime 2}=\alpha (z^2  + 1)$
($\alpha>0$), which is solved to give $z(x)=\pm
\sinh(\sqrt{\alpha}x)$. For definiteness we shall take
$z(x)=\sinh(\sqrt{\alpha}x)$.  The case corresponding to the
negative sign is simply the mirror image of the present case
(i.e., by taking $x\to -x$). The same applies to the other
sinusoidal coordinates discussed in the rest of the paper.

Putting $\beta=0$ and $\gamma=\alpha$ in (\ref{V-2}), we get the
potential ($-\infty < x < \infty$)
\begin{eqnarray}
V_N&=&\frac{A_2^2}{\alpha}\left(z^2+1\right)
+\left[\frac{(A_0-A_2)^2}{\alpha}
-A_1\left(\frac{A_1}{\alpha}+1\right)\right]\frac{1}{z^2+1}
+\left(A_0-A_2\right)\left(\frac{2A_1}{\alpha}+1\right)\frac{z}{z^2+1}\nonumber\\
&&+A_2\left(\frac{2A_1}{\alpha}-2N-1\right)z -\left[2A_1N- \alpha
N^2 -\frac{A_1^2}{\alpha} -\frac{2A_2}{\alpha}(A_0-A_2) +
2A_2\sum_{k=1}^N z_k\right].
\end{eqnarray}
In terms of $x$, it is
\begin{eqnarray}
V_N (x) &=&\frac{A_2^2}{\alpha}\cosh^2(\sqrt{\alpha} x)
+\left[\frac{(A_0-A_2)^2}{\alpha}
-A_1\left(\frac{A_1}{\alpha}+1\right)\right]\rm{sech}^2(\sqrt{\alpha}
x)\nonumber\\
&&+\left(A_0-A_2\right)\left(\frac{2A_1}{\alpha}+1\right)\tanh(\sqrt{\alpha}
x) \rm{sech}(\sqrt{\alpha} x)
+A_2\left(\frac{2A_1}{\alpha}-2N-1\right)\sinh(\sqrt{\alpha} x)
\label{V-Scarf-II}\\
&&-\left[2A_1N- \alpha N^2 -\frac{A_1^2}{\alpha}
-\frac{2A_2}{\alpha}(A_0-A_2) + 2A_2\sum_{k=1}^N z_k\right].
\nonumber
\end{eqnarray}

The prepotential $W_0$, obtained from (\ref{W0}), is
\begin{eqnarray}
W_0(x)&=&\frac{A_2}{\alpha}z +
\frac{A_1}{2\alpha}\ln\left(z^2+1\right) +
\frac{A_0-A_2}{\alpha}\tan^{-1} z\nonumber\\
&=& \frac{A_2}{\alpha}\sinh(\sqrt{\alpha}x) +
\frac{A_1}{\alpha}\ln\cosh(\sqrt{\alpha}x) +
\frac{A_0-A_2}{\alpha}\tan^{-1}\sinh(\sqrt{\alpha}x).
\end{eqnarray}
Hence the ground state wave function $\psi_0\sim e^{-W_0}$ is
\begin{eqnarray}
\psi_0\sim
e^{-\frac{A_2}{\alpha}\sinh(\sqrt{\alpha}x)}\left(\cosh(\sqrt{\alpha}x)
\right)^{-\frac{A_1}{\alpha}}
e^{\frac{A_2-A_0}{\alpha}\tan^{-1}\sinh(\sqrt{\alpha}x)}.\label{wf0-scarf}
\end{eqnarray}

\subsection{$A_2=0$}

Let us first discuss the exactly solvable case, with $A_2=0$. The
potential becomes
\begin{eqnarray}
V_N &=& \left[\frac{A_0^2}{\alpha}
-A_1\left(\frac{A_1}{\alpha}+1\right)\right]\rm{sech}^2(\sqrt{\alpha}
x)+A_0\left( \frac{2A_1}{\alpha}+1\right)\tanh(\sqrt{\alpha}
x)\rm{sech}(\sqrt{\alpha} x)
\nonumber\\
&& -\left[2A_1N- \alpha N^2 -\frac{A_1^2}{\alpha} \right].
\end{eqnarray}
Suppose all $A_i$'s are real, the potential $V_N$ is just the
Scarf II potential given in \cite{Cooper}.  Recall that in our
approach we have $H_N\phi_N=0$, hence the eigenvalue of $H_N$ with
potential $V_N$ is always zero.  If we define the Scarf II
potential $V(x)$ by the first two $N$-independent terms, then
$V_N(x)=V(x)-E_N$, where $E_N=2A_1N- \alpha N^2 -A_1^2/\alpha$ are
the eigenvalues of $V(x)$.  The corresponding eigenfunctions are
given by (\ref{phi_N}), with $\phi_0$ given by (\ref{wf0-scarf}),
and $p_N(z)$ defined by the roots $z_k$'s of the corresponding
Bethe ansatz equation (\ref{BAE}), which gives the Jacobi
polynomials \cite{Szego}.

If we take instead
\begin{equation}
A_0=-i\frac{d}{2},~~\frac{2A_1}{\alpha}+1=-i\frac{c}{\alpha},
\end{equation}
then the potential is
\begin{eqnarray}
V_N(x)&=&\frac{1}{4\alpha}\left(\alpha ^2 +c^2 - d^2\right){\rm
sech}^2(\sqrt{\alpha} x)-\frac{cd}{2\alpha}\tanh(\sqrt{\alpha}
x){\rm sech}(\sqrt{\alpha} x)\nonumber\\
&&-\left[\frac{c^2}{4\alpha} - \left(N+\frac{1}{2}\right)^2\alpha
-ic\left(N+\frac{1}{2}\right)\right].
\end{eqnarray}
This is the exactly solvable case 1 hyperbolic QNM system
discussed in \cite{ChoHo1}.  The special case where $d=0$ has been
employed in \cite{FM} to study black hole's QNMs.   If we take the
potential to be defined by the first two terms, then the energies
and ground state wave function are
\begin{eqnarray}
E_N=\frac{c^2}{4\alpha}  - \left(N+\frac{1}{2}\right)^2\alpha
-ic\left(N+\frac{1}{2}\right).
\end{eqnarray}
and
\begin{eqnarray}
\psi_0(x)\sim (\cosh\sqrt{\alpha} x)^{(ic+\alpha)/2\alpha}\exp(id
\tan^{-1}(\sinh\sqrt{\alpha}x)/2\alpha).
\end{eqnarray}
Note that $E_N$ is independent of $d$, which is a general feature
of the Scarf-type potentials. Also, the imaginary part is
proportional to $N+1/2$, which is characteristic of black hole
QNMs.

\subsection{$A_2\neq 0$}

If $A_2\neq 0$, then reality of the first term of $V_N$ in
(\ref{V-Scarf-II}) requires that $A_2$ be real, or purely
imaginary.

If $A_2$ is real, then it is clear from the first term of
(\ref{wf0-scarf}) that $\psi_0$, which governs the asymptotic
behaviors of $\psi_N$, is not normalizable on the whole line.
Hence there is no QES model with real energies in this case.

Suppose $A_2$ is purely imaginary, say $A_2=ic\neq 0$ with real
constant $c$.  Then the fourth term of $V_N$ in (\ref{V-Scarf-II})
can be real provided that
\begin{equation}
\frac{2A_1}{\alpha}-2N-1=id,~~d:{\rm real}.
\end{equation}
If $d\neq 0$, then reality of the third term of $V_N$ demands that
$A_0-A_2=\pm [2(N+1)-id]\alpha$.  But then, as can be easily
checked, with these values of $A_i$'s the second term of $V_N$
cannot be real, unless $d=0$.

So we are left with the choice $A_2=ic\neq 0$ and
$2A_1/\alpha-2N-1=0$.   This implies that $A_0-A_2$ must be real
for $V_N$ real.  Let $A_0-A_2=a\alpha$ with real $a$.   The
potential of this system assumes the form
\begin{eqnarray}
V_N (x)&=&-\frac{c^2}{\alpha}\cosh^2(\sqrt{\alpha} x)
+\alpha\left[a^2-\left(N+\frac12\right)\left(N+\frac{3}{2}\right)\rm{sech}^2(\sqrt{\alpha}
x)\right]+2a\alpha \left(N+1\right)\tanh(\sqrt{\alpha} x){\rm
sech}(\sqrt{\alpha} x)\nonumber\\
&& -\left(- \frac{\alpha}{4}-2ica + 2ic\sum_{k=1}^N z_k\right).
\label{V-Scarf-new}
\end{eqnarray}
The ground state wave function is
\begin{eqnarray}
\psi_0\sim
e^{-i\frac{c}{\alpha}\sinh(\sqrt{\alpha}x)}\left(\cosh(\sqrt{\alpha}x)
\right)^{-(N+\frac12)} e^{-a\tan^{-1}\sinh(\sqrt{\alpha}x)}.
\end{eqnarray}

For $a\neq 0$, the potential (\ref{V-Scarf-new}) is a new QES
model with QNMs which has been overlooked in our previous study
based on the Lie-algebraic theory of \cite{ChoHo1}.

The case $a=0$ leads to a totally different model with real QES
energies, which was first discussed in \cite{ChoHo2}.  We briefly
discuss it in the next subsection.

\subsection{Singular potential with $A_2\neq 0$}

For $a=0$, i.e., $A_2=A_0=ic\neq 0$, the potential
(\ref{V-Scarf-new}) becomes
\begin{eqnarray}
V_N (x)=-\frac{c^2}{\alpha}\cosh^2(\sqrt{\alpha} x)
-\alpha\left(N+\frac12\right)\left(N+\frac{3}{2}\right)\rm{sech}^2(\sqrt{\alpha}
x) -\left(- \frac{\alpha}{4} + 2ic\sum_{k=1}^N z_k\right).
\end{eqnarray}
This is a singular potential unbounded from below. Yet it exhibits
very peculiar features, such as the existence of QES bound states
with real energies, and QES total transmission modes.  We refer
the reader to \cite{ChoHo2} for a detailed discussion of this
potential.

At first sight it may seem strange that the system has QES real
energies, owing to the term $2ic\sum_k z_k$.  We prove below that
this term is indeed real, as the sum $\sum_k z_k$ is purely
imaginary.

The Bethe ansatz equations (\ref{BAE}) in this case are
\begin{equation}
ic z_k^2 + \alpha N z_k + ic - \alpha \sum_{l\neq k}\frac{ z_k^2 +
1}{z_k-z_l}=0,~~ k=1,2,\ldots,N.\label{BAE-1}
\end{equation}
Multiplying (\ref{BAE-1}) by $-1$ and taking its complex
conjugate, we have
\begin{equation}
ic (-\bar{z}_k)^2 + \alpha N (-\bar{z}_k) + ic - \alpha
\sum_{l\neq k}\frac{ (-\bar{z}_k)^2 +
1}{(-\bar{z}_k)-(-\bar{z}_l)}=0, ~~k=1,2,\ldots,N.\label{BAE-2}
\end{equation}
Here $\bar{z}$ represent the complex conjugate of $z$. Comparing
(\ref{BAE-1}) and (\ref{BAE-2}), we conclude that if $z_k$'s are
the solutions of (\ref{BAE-1}), then so are their negative complex
conjugates $-\bar{z}_k$'s. This implies that the sum $\sum_k z_k$
is purely imaginary, and hence the QES energies are real.

\section{Morse: $z^{\prime 2}=\alpha z^2$ ($\alpha>0$)}

In this case the change of variables is given by
$z(x)=\exp(\pm\sqrt{\alpha} x)$. Again we shall only consider the
positive case, i.e., we take $z(x)=\exp(\sqrt{\alpha} x)$.  The
potential is ($-\infty < x <\infty$)
\begin{eqnarray}
V_N (x) &=& \frac{A_2^2}{\alpha}z^2
+A_2\left(\frac{2A_1}{\alpha}-2N-1\right)z +
A_0\left(\frac{2A_1}{\alpha}+1\right)\frac{1}{z} +
\frac{A_0^2}{\alpha}\frac{1}{z^2} -\left( 2A_1 N - \alpha N^2 -
\frac{A_1^2}{\alpha} - \frac{2A_2A_0}{\alpha}+ 2A_2\sum_{k=1}^N
z_k\right)\nonumber\\
&=&\frac{A_2^2}{\alpha}e^{2\sqrt{\alpha} x}
+A_2\left(\frac{2A_1}{\alpha}-2N-1\right)e^{\sqrt{\alpha} x} +
A_0\left(\frac{2A_1}{\alpha}+1\right)e^{-\sqrt{\alpha} x} +
\frac{A_0^2}{\alpha}e^{-2\sqrt{\alpha} x}\nonumber\\
&& -\left( 2A_1 N - \alpha N^2 - \frac{A_1^2}{\alpha} -
\frac{2A_2A_0}{\alpha}+ 2A_2\sum_{k=1}^N z_k\right).
\label{V-Morse}
\end{eqnarray}
The prepotential $W_0$ is
\begin{eqnarray} W_0 &=&
\frac{1}{\alpha}\left(A_2z + A_1\ln\,z
-\frac{A_0}{z}\right)\nonumber\\
&=& \frac{A_2}{\alpha}e^{\sqrt{\alpha} x} +
\frac{A_1}{\alpha}\sqrt{\alpha} x -
\frac{A_0}{\alpha}e^{-\sqrt{\alpha} x}.
\end{eqnarray}
Hence the ground state wave function is
\begin{eqnarray}
\phi_0\sim e^{-\frac{A_2}{\alpha}e^{\sqrt{\alpha} x} -
\frac{A_1}{\alpha}\sqrt{\alpha} x +
\frac{A_0}{\alpha}e^{-\sqrt{\alpha} x}}.
\end{eqnarray}

\subsection{$A_2=0$}

For $A_2=0$, the potential reads
\begin{eqnarray}
V_N(x)=A_0\left(\frac{2A_1}{\alpha}+1\right)e^{-\sqrt{\alpha} x} +
\frac{A_0^2}{\alpha}e^{-2\sqrt{\alpha} x} -\left( 2A_1 N - \alpha
N^2 - \frac{A_1^2}{\alpha}\right).
\end{eqnarray}
To get a real potential defined by the first two terms of $V_N$,
$A_0$ and $A_1$ must be both real, or both complex.

If $A_0$ is real, then the first term of $V_N$ requires that $A_1$
be real.  The resulted model is the exactly solvable Morse
potential.

If $A_0$ is complex, then the second term of $V_N$ demands that
$A_0$ be purely imaginary, i.e., $A_0=ic$ with $c$ real. The first
term of $V_N$ then requires that $2A_1/\alpha+1=-id$ is also
purely imaginary.  The potential is
\begin{eqnarray}
V_N(x)=cd e^{-\sqrt{\alpha} x} -
\frac{c^2}{\alpha}e^{-2\sqrt{\alpha} x}
-\alpha\left[\frac{d^2-1}{4}-N^2-N
-id\left(N+\frac12\right)\right], \label{V-Morse1}
\end{eqnarray}
with ground state
\begin{equation}
\psi_0\sim
e^{i\frac{c}{\alpha}e^{-\sqrt{\alpha}x}+\frac12\left(id+1\right)
\sqrt{\alpha}x}.
\end{equation}
 This gives a new exactly solvable QNM model.
This system was overlooked in \cite{ChoHo1}.

We note here that this system, or rather its mirror image, can be
obtained by a different complexification of $A_0,~A_1$ and $A_2$.
We shall discuss this in the next subsection.

\subsection{$A_2\neq 0$}

If $A_2\neq 0$, then $A_2$ has to be real or purely imaginary.

If $A_2$ is real, then all $A_i$'s must be real.  In this case,
for $A_2>0$ and $A_0<0$, the potential $V_N$ in (\ref{V-Morse})
defines a QES system.

For $A_2$ purely imaginary, $A_1/\alpha -(N+1/2)$ must either be
zero, or purely imaginary. If $A_1/\alpha -(N+1/2)=0$, we can have
\begin{eqnarray}
A_2=-i\frac{b}{2},~~A_1=\left(N+\frac12\right)\alpha,~~A_0=-\frac{d}{2},~~b,d:
{\rm real}.
\end{eqnarray}
This leads to a potential
\begin{eqnarray}
V_N(x)=-\frac{b^2}{4\alpha} e^{2\sqrt{\alpha} x}- \left(N+1\right)
d e^{-\sqrt{\alpha} x} + \frac{d^2}{4\alpha} e^{-2\sqrt{\alpha}
x}-\left(-  \frac{\alpha}{4}-i\frac{bd}{2\alpha} -ib\sum_{k=1}^N
z_k\right).
\end{eqnarray}
The system defined by the first three terms of $V_N$  is a QES
system with complex energies given by the last term in bracket.
This is indeed the very first QES QNM model presented in
\cite{ChoHo1}.

On the other hand, if $A_1/\alpha -(N+1/2)$ is purely imaginary,
then one can choose
\begin{equation}
A_2=-ic,~~\frac{A_1}{\alpha}=i\frac{d}{2}+\left(N+\frac12\right),~~A_0=0,
\end{equation}
with $c$ and $d$ real constants. The resulted potential is
\begin{eqnarray}
V_N(x)=cd e^{\sqrt{\alpha} x} -
\frac{c^2}{\alpha}e^{2\sqrt{\alpha} x}
-\alpha\left[\frac{d^2-1}{4}
-i\frac{d}{2}-2i\frac{c}{\alpha}\sum_k z_k\right].
\label{V-Morse2}
\end{eqnarray}
In this case though $A_2\neq 0$, the parameter $N$ and the roots
$z_k$'s do not appear in any $x$-dependent term, and hence the
system is exactly solvable.

This system is indeed the mirror image of the model described by
(\ref{V-Morse1}), although the functional forms of the two
potentials look very differently.  To show this, we need to
demonstrate that the potential, the eigenvalues and the
eigenfunctions of one system are mapped into those of the other
system under parity transformation $x\to -x$.

The wave function of the present system is
\begin{eqnarray}
\psi_N\sim
e^{i\frac{c}{\alpha}e^{\sqrt{\alpha}x}-\frac12\left(id+2N+1\right)
\sqrt{\alpha}x}\,\prod_{k=1}^N\,\left(z-z_k\right),
\label{wf-Morse2}
\end{eqnarray}
where the roots $z_k$'s satisfy the BAE (from (\ref{BAE}))
\begin{eqnarray}
-icz_k^2 +\alpha \left(i\frac{d}{2}+N\right)z_k -\alpha
\sum_{l\neq k}^N \frac{z_k^2}{z_k-z_l}=0,~~ k=1,2,\ldots,N.
\label{BAE-Morse2}
\end{eqnarray}
We stress here that $z$ is $z=e^{\sqrt{\alpha}x}$ as before.

Under parity transformation $x\to -x$, we have $z\to
z^{-1},~z_k\to z_k^{-1}$.  Eqs.~(\ref{V-Morse2}),
(\ref{wf-Morse2}) and (\ref{BAE-Morse2}) are mapped, respectively,
into
\begin{eqnarray}
V_N(x)&=&cd e^{-\sqrt{\alpha} x} -
\frac{c^2}{\alpha}e^{-2\sqrt{\alpha} x}
-\alpha\left[\frac{d^2-1}{4}
-i\frac{d}{2}-2i\frac{c}{\alpha}\sum_k \frac{1}{z_k}\right],
\label{V-Morse2-1}\\
\psi_N &\sim &
e^{i\frac{c}{\alpha}e^{-\sqrt{\alpha}x}+\frac12\left(id+2N+1\right)
\sqrt{\alpha}x}\,\prod_{k=1}^N\,\left(\frac{1}{z}-\frac{1}{z_k}\right),
\label{wf-Morse2-1}
\end{eqnarray}
and
\begin{eqnarray}
-ic\frac{1}{z_k^2} +\alpha
\left(i\frac{d}{2}+N\right)\frac{1}{z_k} -\alpha \sum_{l\neq k}^N
\frac{z_l}{z_k(z_l-z_k)}=0,~~ k=1,2,\ldots,N. \label{BAE-Morse2-1}
\end{eqnarray}

Multiplying (\ref{BAE-Morse2-1}) by $-z_k^2$, and using the
identity
\begin{eqnarray}
\sum_{l\neq k}^N \frac{z_l}{z_l-z_k}=N-1-\sum_{l\neq k}^N
\frac{z_k}{z_k-z_l},
\end{eqnarray}
we can transform (\ref{BAE-Morse2-1}) to
\begin{eqnarray}
-\alpha\left(i\frac{d}{2}+1\right)z_k +ic -\alpha \sum_{l\neq k}^N
\frac{z_k^2}{z_k-z_l}=0, ~~k=1,2,\ldots,N. \label{BAE-Morse2-2}
\end{eqnarray}
This is just the BAE for the roots $z_k$'s of the eigenfunctions
of the system defined by the potential (\ref{V-Morse1}). Hence the
BAE of this system is mapped into the BAE of the system defined by
(\ref{V-Morse1}) under parity.

Now the eigenfunctions (\ref{wf-Morse2-1}) can be rewritten as
\begin{eqnarray}
\psi_N &\sim &
e^{i\frac{c}{\alpha}e^{-\sqrt{\alpha}x}+\frac12\left(id+1\right)
\sqrt{\alpha}x}\,\left(e^{\sqrt{\alpha}x}\right)^N\,z^{-N}\prod_{k=1}^N
\,\left(z-z_k\right),\nonumber\\
&\sim &
e^{i\frac{c}{\alpha}e^{-\sqrt{\alpha}x}+\frac12\left(id+1\right)
\sqrt{\alpha}x}\,\prod_{k=1}^N\,\left(z-z_k\right),
\label{wf-Morse2-2}
\end{eqnarray}
where we have used the fact that $(e^{\sqrt{\alpha}x})^N
z^{-N}=1$. Eq.~(\ref{wf-Morse2-2}) is just the eigenfunction of
the potential (\ref{V-Morse1}).  Thus, together with the result of
the last paragraph on BAE, one sees that under parity the
eigenfunctions of the system given by (\ref{V-Morse2}) are mapped
into those of given by (\ref{V-Morse1}).

Finally we show that the last term of the potential
(\ref{V-Morse2}) is equal to the last term of (\ref{V-Morse1}).
This  proves the equality of the eigenvalues of both systems.
Dividing the BAE (\ref{BAE-Morse2-2}) by $z_k$, summing over all
$k$ and using the identity
\begin{eqnarray}
\sum_{k=1}^N\sum_{l\neq k} \frac{z_k}{z_k-z_l}=\frac12 N(N-1),
\end{eqnarray}
we get
\begin{equation}
-2ic\frac{c}{\alpha}\sum_{k=1}^N\frac{1}{z_k}=-id N -N^2-N.
\end{equation}
So the last term of the potential (\ref{V-Morse2}) becomes
\begin{eqnarray}
-\alpha\left[\frac{d^2-1}{4}
-i\frac{d}{2}-2i\frac{c}{\alpha}\sum_k z_k\right] =-\alpha\left[
\frac{d^2-1}{4}-N^2-N -id\left(N+\frac12\right) \right].
\end{eqnarray}
This is equal to the last term of (\ref{V-Morse1}).  Hence under
parity potential (\ref{V-Morse2}) is mapped into (\ref{V-Morse1}).

\section{Generalized P\"oschl-Teller: $z^{\prime 2}=\alpha (z^2  - 1)$}

In this case $z(x)=\cosh(\sqrt{\alpha}x)$. The potential is
($0\leq x<\infty$)
\begin{eqnarray}
V_N (x) &=&\frac{A_2^2}{\alpha}\sinh^2(\sqrt{\alpha} x)
+\left[\frac{(A_0+A_2)^2}{\alpha}
+A_1\left(\frac{A_1}{\alpha}+1\right)\right]\rm{cosech}^2(\sqrt{\alpha}
x)\nonumber\\
&&+\left(A_0+A_2\right)\left(\frac{2A_1}{\alpha}+1\right)
\coth(\sqrt{\alpha} x)\rm{cosech}(\sqrt{\alpha} x)
+A_2\left(\frac{2A_1}{\alpha}-2N-1\right)\cosh(\sqrt{\alpha}
x)\label{V-PT} \\
&& -\left[2A_1N- \alpha N^2 -\frac{A_1^2}{\alpha}
-\frac{2A_2}{\alpha}(A_0+A_2) + 2A_2\sum_{k=1}^N z_k\right].
\nonumber
\end{eqnarray}
The ground state wave function $\psi_0\sim\exp(-W_0)$ is
\begin{eqnarray}
\psi_0&\sim& e^{-\frac{A_2}{\alpha}\cosh(\sqrt{\alpha}x)}
\left(\sinh(\sqrt{\alpha}x)\right)^{-\frac{A_1}{\alpha}}
e^{\frac{A_2+A_0}{\alpha}\coth^{-1}\cosh(\sqrt{\alpha}x)}\nonumber\\
&\sim &e^{-\frac{A_2}{\alpha}\cosh(\sqrt{\alpha}x)}
\left(\sinh(\sqrt{\alpha}x)\right)^{-\frac{A_1}{\alpha}}
\left[\tanh(\frac{\sqrt{\alpha}}{2}
x)\right]^{-\frac{A_2+A_0}{\alpha}}.
\end{eqnarray}
Since $V_N\to \infty $ as $x\to 0$, we must have the boundary
condition $\psi_N\to 0$ as $x\to 0$.

\subsection{$A_2=0$}

For $A_2=0$, the potential is
\begin{eqnarray}
V_N &=& \left[\frac{A_0^2}{\alpha}
+A_1\left(\frac{A_1}{\alpha}+1\right)\right]\rm{cosech}^2(\sqrt{\alpha}
x)+A_0\left(\frac{2A_1}{\alpha}+1\right)\coth(\sqrt{\alpha} x)
\rm{cosech}(\sqrt{\alpha} x)\nonumber\\
&& -\left[2A_1N- \alpha N^2 -\frac{A_1^2}{\alpha} \right].
\end{eqnarray}
If all $A_i$'s are real, this gives the generalized
P\"oschl-Teller potential in \cite{Cooper}.  If we take the
P\"oschl-Teller potential $V(x)$ to consist of the first two
$N$-independent terms, then the eigen-energies are $E_N=2A_1N-
\alpha N^2 -A_1^2/\alpha$.  These eigenvalues are the same as
those of the Scarf II potential. Our approach makes it very easy
to see why the eigenvalues are the same for the two apparently
different systems.

We can also take
\begin{equation}
A_0=-i\frac{d}{2},~~\frac{2A_1}{\alpha}+1=-i\frac{c}{\alpha},
\end{equation}
which result in the potential
\begin{eqnarray}
V_N(x)&=&-\frac{1}{4\alpha}\left(\alpha ^2 +c^2 + d^2\right){\rm
cosech}^2(\sqrt{\alpha} x)-\frac{cd}{2\alpha}\coth(\sqrt{\alpha}
x){\rm cosech}(\sqrt{\alpha} x)\nonumber\\
&&-\left[\frac{c^2}{4\alpha} - \left(N+\frac{1}{2}\right)^2\alpha
-ic\left(N+\frac{1}{2}\right)\right].
\end{eqnarray}
This is the exactly solvable case 2 hyperbolic QNM system
discussed in \cite{ChoHo1}.

\subsection{$A_2\neq 0$}

As with the Scarf II case, if $A_2\neq 0$, then reality of the
first term of $V_N$ in (\ref{V-PT}) requires that $A_2$ be real,
or purely imaginary.

If $A_2$ is real, then $A_0$ and $A_1$ have to be real. The
boundary condition that $\psi_N\to 0$ as $x\to \infty$ is met if
$A_2>0$.  As $x\to 0$, we have
\begin{equation}
\psi_0\sim x^{-\frac{A_2+A_1+A_0}{\alpha}}.
\end{equation}
Thus $\psi_0\to 0$ as $x\to 0$ is guaranteed if
\begin{equation}
 A_2+A_1+A_0<0.\label{A-cond}
\end{equation}
With $A_i$'s satisfying (\ref{A-cond}) and $A_2>0$, the potential
$V_N$ gives a QES system with real energies.

If $A_2$ is imaginary, say $A_2=ic\neq 0$ with real constant $c$,
then the fourth term of $V_N$ in (\ref{V-PT}) can be real if
$2A_1/\alpha-2N-1=0$ or $id\neq 0$.  As in the Scarf II case, if
$2A_1\alpha-2N-1=id\neq 0$, then reality of the third term of
$V_N$ demands that $A_0+A_2=\pm [2(N+1)-id]\alpha$.  But then the
second term of $V_N$ cannot be real, unless $d=0$.

So we must have $2A_1/\alpha-2N-1=0$. In this case one has
$A_2+A_0=-a\alpha\neq 0$ with real $a$.  The potential $V_N$ is
\begin{eqnarray}
V_N (x) &=&-\frac{c^2}{\alpha}\sinh^2(\sqrt{\alpha} x)
+\alpha\left[a^2 +
\left(N+\frac12\right)\left(N+\frac{3}{2}\right)\right]\rm{cosech}^2(\sqrt{\alpha}
x)\nonumber\\
&&-2a\alpha \left(N+1\right) \coth(\sqrt{\alpha}
x)\rm{cosech}(\sqrt{\alpha} x)
\label{V-PT-2} \\
&& -\left(- \frac{\alpha}{4}+2ica + 2ic\sum_{k=1}^N z_k\right),
\nonumber
\end{eqnarray}
and the ground state wave function
\begin{eqnarray}
\psi_0&\sim& e^{-i\frac{c}{\alpha}\cosh(\sqrt{\alpha}x)}
\left(\sinh(\sqrt{\alpha}x)\right)^{-(N+\frac12)}
\left[\tanh(\frac{\sqrt{\alpha}}{2} x)\right]^{a}.
\end{eqnarray}
To satisfy the boundary condition $\psi_0\to 0$ as $x\to 0$, we
have from (\ref{A-cond}) the condition
\begin{equation}
a>N+\frac12.
\end{equation}
With these $A_i$'s, we have a new QES system with QNMs.

Unlike the Scarf II case, however, when $a=0$, we do not have a
QES model with real energy, since in this case $\psi_0$ is not
normalizable at $x=0$.

\section{Summary}

Exactly solvable models admitting QNM maybe useful in providing
insights to QNMs emitted from more complicated systems such as
black holes.  In \cite{ChoHo1} we have found some new exactly
solvable QNM models, and a new QES QNM model within the $sl(2)$
Lie-algebraic approach to QES theory.

In this paper we have demonstrated how exactly solvable and QES
models admitting quasinormal modes can be constructed and
classified very simply and directly by the prepotential approach.
This approach,  unlike the Lie-algebraic approach, does not
require any knowledge of the underlying symmetry of the system. It
treats both quasi-exact and exact solvabilities on the same
footing, and gives the potential as well as the eigenfunctions and
eigenvalues simultaneously. We also present a new exactly solvable
Morse-like model with quasinormal modes, and two new quasi-exactly
solvable models with quasinormal modes of the Scarf II and
generalized P\"oschl-Teller types.

\begin{acknowledgments}

This work is supported in part by the National Science Council
(NSC) of the Republic of China under Grant NSC
96-2112-M-032-007-MY3 and NSC-99-2112-M-032-002-MY3.

\end{acknowledgments}

\appendix

\section{Shifted oscillator: $z^{\prime 2}=\gamma>0$}

The corresponding transformation $z(x)$ is
$z(x)=\sqrt{\gamma}x+\rm{constant}$. By an appropriate translation
one can always set the constant to zero. Hence we shall take
$z(x)=\sqrt{\gamma}x$ without loss of generality.

The potential is ($-\infty < x < \infty$)
\begin{eqnarray}
V_N &=&A_2^2 \gamma x^4 + 2 A_2A_1 \sqrt{\gamma}x^3 +\left(A_1^2 +
2A_2A_0\right)x^2 + 2\left[\frac{A_1A_0}{\gamma}
-A_2(N+1)\right]\sqrt{\gamma}x\nonumber\\
&&-\left[2A_1(N+\frac12)-\frac{A_0^2}{\gamma}+ 2A_2\sum_{k=1}^N
z_k\right].
\end{eqnarray}
The ground state wave function $\psi_0\sim\exp(-W_0)$ is
\begin{eqnarray}
\psi_0\sim e^{-\frac{A_2}{3}\sqrt{\gamma} x^3 -\frac{A_1}{2}x^2 -
\frac{A_0}{\sqrt{\gamma}}x}.
\end{eqnarray}

We want $V_N$ to be real. If $A_2\neq 0$, then the term $A_2(N+1)$
in the fourth term implies $A_2$ be real. However, this implies
the wave function $\psi_N$, whose asymptotic behavior is governed
by $\psi_0$, is not normalizable on the whole line. Hence there is
no QES models, with or without QNMs.

For $A_2=0$, the potential is
\begin{eqnarray}
V_N=\left(A_1 x  +
\frac{A_0}{\sqrt{\gamma}}\right)^2-2A_1(N+\frac12).
\end{eqnarray}
So $A_1$ and $A_0$ must both be real, this is just the potential
of the shifted oscillator.

If we take
\begin{equation}
A_1=-i\frac{c}{2},~~A_0=0,
\end{equation}
the potential becomes
\begin{equation}
V_N=-\frac{1}{4}c^2x^2 + ic \left(N+\frac12\right).
\end{equation}
This is the case 5 exactly solvable QNM model discussed in
\cite{ChoHo1}.

\section{Radial oscillator: $z^{\prime 2}=\beta z,~~(\beta>0$)}

For simplicity we take $z(x)=\frac{\beta}{4}x^2$.  The potential
is ($0\leq x < \infty$)
\begin{eqnarray}
V_N &=& \frac{1}{64}A_2^2\beta^2 x^6 + \frac{1}{8}A_2A_1\beta x^4
+\frac{1}{4}\left[A_1^2+
2A_2A_0-\frac{1}{2}{A_2\beta}\left(4N+3\right)
\right]x^2\nonumber\\
&&
+\frac{4A_0}{\beta}\left(\frac{A_0}{\beta}+\frac{1}{2}\right)\frac{1}{x^2}
-\left[\frac{A_1}{2}\left(4N+1-\frac{4A_0}{\beta}\right)+2A_2\sum_{k=1}^N
z_k\right]. \label{V-3d-osc}
\end{eqnarray}
The ground state $\psi_0\sim\exp(-W_0)$ is
\begin{equation}
\psi_0\sim
x^{-\frac{2A_0}{\beta}}e^{-\frac{1}{4}A_1x^2-\frac{1}{32}A_2\beta
x^4}.
\end{equation}

It can be checked that if $A_2\neq 0$, then all $A_i$'s have to be
real. Hence there is no QES model with QNMs in this case. For
$A_2>0$ and $A_0<0$, the wave functions $\psi_N$ are normalizable
on the positive half-line. The system is then a QES model with
real energies.  Together with the discussion in Appendix~A, one
concludes that one-dimensional QES models start with degree six,
i.e., the sextic oscillator \cite{note}.   In fact, the potential
(\ref{V-3d-osc}) with $A_2=2a>0,~A_1=2b,~A_0=0$ and $\beta=4$,
namely,
\begin{eqnarray}
V_N&=&a^2 x^6 +2ab x^4+ \left[b^2-(4N+3)a\right]x^2 \nonumber\\
&&-\left[(4N+1)b+4a{\sum_k z_k}\right] ,
\end{eqnarray}
is the very first QES model discussed in the literature \cite{TU}.

If $A_2=0$, then (\ref{V-3d-osc}) becomes
\begin{equation}
V_N=\frac{A_1^2}{4}x^2
+\frac{4A_0}{\beta}\left(\frac{A_0}{\beta}+\frac12\right)\frac{1}{x^2}
- A_1\left(2N+\frac12 -\frac{2A_0}{\beta}\right).
\end{equation}
For real $A_1$ and $ A_0$, this is just the radial oscillator.

Suppose we take
\begin{equation}
\beta=4,~~A_1=-2ia,~~{\rm and}~~A_0=-2\gamma.
\end{equation}
The potential takes the form
\begin{equation}
V_N=-a^2x^2 +\frac{\gamma(\gamma - 1)}{x^2} + ia \left(4N+2\gamma
+1\right).
\end{equation}
This is the Case 4 model with QNMs presented in \cite{ChoHo1}.

\section{Scarf I: $z^{\prime 2}=\alpha (1 - z^2)$}

In this case $z(x)=\sin(\sqrt{\alpha}x)$. The potential is
\begin{eqnarray}
V_N (x) &=&\frac{A_2^2}{\alpha}\cos^2(\sqrt{\alpha} x)
+\left[\frac{(A_0+A_2)^2}{\alpha}
+A_1\left(\frac{A_1}{\alpha}-1\right)\right]\sec^2(\sqrt{\alpha}
x)+\left(A_0+A_2\right)\left(\frac{2A_1}{\alpha}-1\right)
\tan(\sqrt{\alpha} x)\sec(\sqrt{\alpha} x)\nonumber\\
&&-A_2\left(\frac{2A_1}{\alpha}+2N+1\right)\sin(\sqrt{\alpha} x)
-\left[2A_1N+ \alpha N^2 +\frac{A_1^2}{\alpha}
+\frac{2A_2}{\alpha}(A_0+A_2) + 2A_2\sum_{k=1}^N z_k\right],
\end{eqnarray}
with the ground state wave function
\begin{eqnarray}
\psi_0\sim
e^{\frac{A_2}{\alpha}\sin(\sqrt{\alpha}x)}\left(\cos(\sqrt{\alpha}x)
\right)^{\frac{A_1}{\alpha}}
e^{-\frac{A_2+A_0}{\alpha}\tanh^{-1}\sin(\sqrt{\alpha}x)}.\label{Scarf-I-wf}
\end{eqnarray}
The system is defined only on a finite interval, usually taken to
be $\sqrt{\alpha}x\in [-\pi/2,\pi/2]$.  Hence there are no QNMs.

If $A_2\neq 0$, then all $A_i$'s have to be real.  The third term
of (\ref{Scarf-I-wf}) then implies that the wave functions
$\psi_N$ are not normalizable.  So there are no QES systems with
real energies.

For $A_2=0$, the potential becomes
\begin{eqnarray}
V_N &=& \left[\frac{A_0^2}{\alpha}
+A_1\left(\frac{A_1}{\alpha}-1\right)\right]\sec^2(\sqrt{\alpha}
x)+A_0\left(\frac{2A_1}{\alpha}-1\right)\tan(\sqrt{\alpha} x)
\sec(\sqrt{\alpha} x)\nonumber\\
&& - \left[2A_1N + \alpha N^2 + \frac{A_1^2}{\alpha} \right].
\end{eqnarray}
For real $A_1$ and $A_0$, this is Scarf I potential given in
\cite{Cooper}.

% --------------------------------

\end{document}